\documentclass{iaus}
\usepackage{graphicx}

\def\a4{\hsize 17.0cm \vsize 25.cm}

\title[On the structure of giant HII regions]
{On the structure of giant HII regions and HII galaxies}

\author[Tenorio-Tagle et al.] 
{Tenorio-Tagle, G.$^1$ \break
C. Mu\~noz-Tu\~n\'on $^2$  
E. P\'erez$^3$  
S. Silich$^1$ 
E. Telles$^4$}
\affiliation{$^1$ Instituto Nacional de Astrof\'\i sica Optica y
Electr\'onica, AP 51, 72000 Puebla, M\'exico \break email: gtt@inaoep.mx
\\[\affilskip]
$^2$ Instituto de Astrof\'{\i}sica de Canarias, E 38200 La
Laguna, Tenerife, Spain \break cmt@ll.iac.es
\\[\affilskip]
$^3$ Instituto de Astrof\'{\i}sica de Andaluc\'\i{}a (CSIC), 
Camino bajo de Huetor 50, E 18080 Granada, Spain  \break eperez@iaa.es
\\[\affilskip]
$^4$ Observat\'orio Nacional, 
Rua Jos\'e Cristino 77, 20921-400, Rio de Janeiro, Brazil \break etelles@on.br
}

\pubyear{2006}
\volume{237}  %% insert here IAU Symposium No.
\pagerange{0--4}
\date{?? and in revised form ??}
\jname{Triggered Star Formation in a Turbulent ISM}
\editors{B.G. Elmegreen \& J.  Palou\v{s}, eds.}
\begin{document}

\maketitle

\begin{abstract}
We review the structural properties of giant extragalactic HII regions and HII galaxies  
based on  two dimensional hydrodynamic calculations, and
propose an evolutionary sequence that accounts for their observed detailed structure.
The model assumes a massive and young stellar cluster surrounded by a large collection of clouds. These are thus 
exposed to the  most important star-formation feedback mechanisms:  photoionization and the cluster wind.  
The models show how the two feedback mechanisms compete with each other in the disruption of clouds 
and lead to two different hydrodynamic solutions: The storage of clouds into a long lasting ragged shell that inhibits the
expansion of the thermalized wind, and the steady filtering of the shocked wind gas through channels 
carved within the cloud stratum that results into the creation of large-scale superbubbles. 
Both solutions are here claimed to be concurrently at work in giant HII regions and HII galaxies, 
causing their detailed inner structure. 
{\bf A full description of the calculations can be found in The Astrophysical Journal 
643, 186. Animated version of the models presented can be 
found at http://www.iaa.csic.es/\~{}eperez/ssc/ssc.html.}
\end{abstract}

\section{Introduction}

Multiple studies during the last decades have addressed the impact of photoionization, stellar winds and supernova explosions 
on interstellar matter (ISM). Disruptive events that re-structure the birth place of massive stars and their surroundings,
while leading to multiple  phase transitions in the ISM (see e.g. Comeron 1997; Yorke et al. 1989; 
Garc\'i{}a-Segura et al. 2004; Hosokawa \& Inutsuka 2005 and  Tenorio-Tagle \& Bodenheimer 1988 and references therein). On the other hand,
giant molecular clouds, the sites of ongoing star formation, present a hierarchy of clumps and filaments of different scales
whose volume filling factor varies between  10$\%$ to 0.1$\%$ (see McLow \& Klessen, 2004, and references therein). 
As the efficiency of star formation in molecular clouds is estimated to be $\leq $ 10$\%$ (Larson 1988; Franco et al. 1994) the implication is that 
the bulk of the cloud structure remains after a star forming episode.
All of these studies are relevant within the fields of interstellar matter and star formation 
and, in particular, regarding  the physics of feedback, a major ingredient in the evolution of galaxies and thus in cosmology.

From the observations, we know that 
giant extragalactic HII regions and HII galaxies are excellent examples of the impact of massive stars on the ISM. They  
belong to the same class of objects because  they are powered by 
young  massive bursts of star formation. Also because of their similar physical size, morphology and inner structure. 
Detailed studies of 30 Doradus (see Chu \& Kennicutt 1994; Melnick et al. 1999), NGC 604 (Sabalisck et al. 1995; Yang et al 1996) and 
several giant extragalactic HII regions and HII galaxies (Mu\~noz-Tu\~n\'on et al. 1996; 
Telles et al. 2001; Ma\'\i z-Apell\'aniz et al. 1999)
have been designed with the aim of unveiling the inner structure and dynamics of the nearest 
examples.
All of them present a collection of nested shells that enclose an X-ray emmiting gas and that 
may extend up to kpc scales. Some of the largest shells 
have stalled while others present 
expansion speeds of up to several tens of km s$^{-1}$. Detailed HST images
have also confirmed these issues  for HII galaxies. 
All of these sources, as in the studies of Telles et al (2001) and Cair\'os et al. (2001), present  
a central  bright condensation coincident with the massive burst of stellar formation. 
In giant HII regions and in some HII galaxies, this is resolved as the brightest
filament or a broken ragged shell sitting very close to the exciting cluster.

\section{Feedback from massive star clusters}

The energy powering these giant volumes comes from the recently found unit of massive star formation: 
super star clusters (Ho 1997).
Massive concentrations of young stars within the range of 10$^5$ M$_\odot$ to several $10^6$ M$_\odot$, 
all within a small volume $\sim$ 3 -- 10 pc.
The mechanical energy and the UV photon output from a massive stellar cluster is here confronted with an ISM
structured into a collection of clouds. 

There are two competing events:
The pressure acquired by the wind at the reverse shock ($P_{bubble} = \rho_w v_{\infty}^2$) 
defines the velocity that the leading shock
may have as it propagates into the intercloud medium ($V_S = (P_{bubble} / \rho_{ic})^{0.5}$), 
and thus, as a first approximation, if the cloudlet distribution has an extent 
$D_{cl}$ the time for the leading shock to travel across it is $t_{s}$ ($= D_{cl} / V_S$). On the other hand, 
the pressure gradient between ionized cloudlets and the intercloud medium, established through photoionization,
is to disrupt clouds and lead to a constant density medium in a time $t_{d} = \alpha d_{cs} / (2 c_{HII})$;
where $d_{cs}/2$ is half the average distance or separation between 
clouds, $c_{HII}$  is the sound velocity in the ionized medium 
and $\alpha$ is a small number ($\sim$ 4 - 6) and accounts for the number of times that a 
rarefaction wave ought to travel (at the sound speed) the 
distance $d_{cs}/2$ to replenish the whole volume with an average even
density $<\rho>$. In this way if the cloud disruption process 
promoted by photoionization leads to an average $<\rho>$ that 
could largely reduce the velocity of the leading shock 
($V_S = (P_{bubble} / <\rho>)^{0.5}$), then $t_{s}$ could be larger 
than  $t_{d}$, leading, as shown below, to an effective
confinement of the shocked wind, at least for a significant part of the evolution.

Cases considering both feedback events lead to a  rapid evolution of the photoionized cloud stratum,
filling almost everywhere the low density intercloud zones, while 
spreading the density of the outermost cloudlets into the surrounding intercloud medium.
This causes the development of an increasingly larger ionized expanding  rim around the cloudlet distribution
and  leads to a rapid enhancement of the intercloud density. 
On the other hand, the wind - cloudlet interaction  leads to a 
global reverse shock into the wind, that evolves from  an initially ragged surface across which the 
isotropic wind is only partly thermalized, to an almost hemispherical surface that fully thermalizes the wind. 
Initially, after crossing the reverse shock
the hot wind drives the leading shock into the cloudlet distribution and this immediately looks for
all possible paths of least resistance in between clouds. This fact, 
diverts the shocked wind into multiple streams behind every overtaken cloudlet,   
diminishing steadily its power  
to reach the end of the cloudlet distribution before the outermost clouds expand and block many of the possible exits
into the low density background gas. 

In the calculation only one channel, the initially widest channel close to the grid equatorial axis, 
across which the original cloudlet spacing was set 
slightly larger than in the rest of the distribution, is succesfully crossed by the 
leading shock and the thermalized wind behind it, after $\sim$ 10$^5$ yr of evolution. 
A second channel through the cloud stratum is completed 
(close to the symmetry axis) after a time 
$t = 5 \times 10^5$ yr. 
All other possibilities,  are blocked by the 
large densities resultant from the champagne bath.

\section{Discussion}

%---------------------------------------------------------------

We have shown here the effects of feedback from a  massive stellar cluster into a selected cloudlet 
distribution, which although arbitrary, as it could have had a different extent 
or it could have considered clouds of different sizes,
separations and locations, it has allowed us to explore a wide range of possibilities. These however, have lead  
to the two main 
possible physical solutions to be expected: partial pressure confinement of the shocked wind, 
and a stable and long-lasting filtering of the thermalized wind 
through the cloudlet stratum. 
Thus despite the arbitrary and simple boundary and initial conditions  
here assumed, the structures that develop within the flow resemble  the structure of
giant HII regions and HII galaxies.
In particular we refer to the giant and multiple well structured shells evident at optical wavelengths, often referred to as nested shells 
(see Chu \& Kenniccutt 1994), 
that enclose a hot X-ray emitting gas, as in 30 Dor (Wang 1999) and NGC 604 (Ma\'\i{}z-Apell\'aniz et al. 2004). 
The calculations also lead to the slowly expanding brightest filament (or ragged shell) 
present in all sources close and around the exciting stars, and to the elongated, 
although much fainter,  filaments on either side of the 
open channels that reseamble the ends of elongated columns into the giant superbubbles.
 
None of these structural features appear in calculations 
that assume a constant density ISM, nor in those that allow for a constant density molecular cloud as birth
place of the exciting sources, or in calculations where the sources are embedded in a plane stratified background atmosphere.
For all of these features to appear, a clumpy circumstellar medium seems to be a necessary requirement. A medium that would
refrain the wind from an immediate exit into the surrounding gas and that 
would also allow  for the build up of multiple channels through 
which the wind energy would flow in a less unimpedded manner. And thus 
giant HII regions and HII galaxies, both powered by massive star formation events producing  an ample supply of UV 
photons and a powerful wind mechanical energy, all of them seem to process their energy into a stratum of dense cloudlets 
sitting in the immediate vicinity of the star formation event and this leads to the repeated structure common to all giant
nebulae.

\acknowledgments
This study has been 
partly supported by grants AYA2004-08260-C03-01 and AYA 2004-02703
from the Spanish Ministerio de Educaci\'on y Ciencia, grant TIC-114 
from Junta de Andaluc\'\i a and Conacyt (M\'exico) grant 47534-F.

\begin{discussion}

\discuss{B. Elmegreen}
{What kind of coherent resistance to the wind will an interconnecting 
magnetic field give if the field connects the clouds with the intercloud 
medium?}

\discuss{G. Tenorio-Tagle}
{A large magnetic pressure would enhance the resistance that the ragged 
shell presents to the shocked wind. In such cases I expect the ragged shell 
to have a longer life time despite its nearness to the exciting sources.}

\discuss{S. Shore} 
{Your calculational scenario has in effect created a "fluffy screen" so your 
flow is more or less a supersonic grid turbulence (without the ionization 
effects) so: what is the effective Reynolds number for the large scale? How 
do your results change with changes in the resolution?
Have you tried to use a random and/or fractal structure for the surrounding 
medium? You might compare your results with laboratory studies.}

\discuss{G. Tenorio-Tagle}
{I am sorry but I don't understand what do you mean by a "fluffy screen". 
We have included in our calculations a realistic computational representation 
of the feedback from super stellar clusters. Many test calculations reassure 
us that the results are independent of the adopted numerical resolution. We 
would like indeed to follow your suggestion and surround our clusters 
by a random fractal structure. However, I would expect that if a significant 
degree of clumpiness is adopted as initial condition, then the two solutions 
here described would be recovered.} 

\discuss{J. Turner}
{Have you included the effects of gravity in these models? For SSCs, 
this could be an important effect, and could, like in stars,
facilitate a smooth transition to supersonic flow.}

\discuss{G. Tenorio-Tagle}
{Gravity is not relevant to super stellar clusters able to drive a stationary 
wind. This is because of the efficient thermalization of the deposited matter 
which leads to large temperatures ($T \sim 10^7$ K) and thus to 
sound velocities that well exceed the escape velocity.}

\end{discussion}

\end{document}